\documentclass[aps,prb,floatfix,twocolumn,superscriptaddress,showpacs]{revtex4-1}
\usepackage{graphicx}
\usepackage{epstopdf}
\usepackage[normalem]{ulem}
\usepackage{amsmath}
\usepackage{pstricks}
\usepackage{color}

\newcommand{\simlt}
      {\ifmmode       { \raisebox{-.8em}{$<$}\atop\sim}
         \else        {$\raisebox{-.8em}{$<$}\atop\sim$}
      \fi}

\begin{document}
\title{Impact of Iron-site defects on Superconductivity in LiFeAs}
\author{Shun Chi}
\altaffiliation{These authors contributed equally.}
\affiliation{Department of Physics and Astronomy, University of British Columbia, Vancouver BC, Canada V6T 1Z1}
\affiliation{Quantum Matter Institute, University of British Columbia, Vancouver BC, Canada V6T 1Z4}
\author{Ramakrishna Aluru}
\altaffiliation{These authors contributed equally.}
\affiliation{Max-Planck-Institut f\"ur Festk\"orperforschung, Heisenbergstr. 1, D-70569 Stuttgart, Germany}
\affiliation{SUPA, School of Physics and Astronomy, University of St. Andrews, North Haugh, St. Andrews, Fife, KY16 9SS, United Kingdom}
\author{Udai Raj Singh}
\affiliation{Max-Planck-Institut f\"ur Festk\"orperforschung, Heisenbergstr. 1, D-70569 Stuttgart, Germany}
\author{Ruixing Liang}
\affiliation{Department of Physics and Astronomy, University of British Columbia, Vancouver BC, Canada V6T 1Z1}
\affiliation{Quantum Matter Institute, University of British Columbia, Vancouver BC, Canada V6T 1Z4}
\author{Walter N. Hardy}
\affiliation{Department of Physics and Astronomy, University of British Columbia, Vancouver BC, Canada V6T 1Z1}
\affiliation{Quantum Matter Institute, University of British Columbia, Vancouver BC, Canada V6T 1Z4}
\author{D. A. Bonn}
\affiliation{Department of Physics and Astronomy, University of British Columbia, Vancouver BC, Canada V6T 1Z1}
\affiliation{Quantum Matter Institute, University of British Columbia, Vancouver BC, Canada V6T 1Z4}
\author{A. Kreisel}
\email{kreisel@itp.uni-frankfurt.de}
\affiliation{Niels Bohr Institute, University of Copenhagen, Universitetsparken 5, DK-2100 Copenhagen,
Denmark}
\author{Brian M. Andersen}
\affiliation{Niels Bohr Institute, University of Copenhagen, Universitetsparken 5, DK-2100 Copenhagen,
Denmark}
\author{R. Nelson}
\affiliation{Dept. of Physics and Astronomy, Louisiana State University,Baton Rouge, LA 70803 USA }
\author{T. Berlijn}
\affiliation{Center for Nanophase Materials Sciences, Oak Ridge National Laboratory, Oak Ridge, Tennessee 37831, USA}
\affiliation{Computer Science and Mathematics Division, Oak Ridge National Laboratory, Oak Ridge, Tennessee 37831, USA}
\author{W. Ku}
\affiliation{CMPMSD, Brookhaven National Laboratory, Upton, NY 11973 USA}
\author{P. J. Hirschfeld}
\affiliation{Dept. of Physics, U. Florida, Gainesville, FL 32611 USA}
\author{Peter Wahl}
\email{wahl@st-andrews.ac.uk}
\affiliation{SUPA, School of Physics and Astronomy, University of St. Andrews, North Haugh, St. Andrews, Fife, KY16 9SS, United Kingdom}
\affiliation{Max-Planck-Institut f\"ur Festk\"orperforschung, Heisenbergstr. 1, D-70569 Stuttgart, Germany}

\date{\today}

\begin{abstract}

In conventional  $s$-wave superconductors, only magnetic impurities exhibit impurity bound states, whereas    for an $s_\pm$ order parameter  they can occur  for both magnetic and non-magnetic impurities. Impurity bound states in superconductors can thus provide important insight into the order parameter. Here, we present a combined experimental and theoretical study of native and engineered iron-site defects in LiFeAs.
Detailed comparison of tunneling spectra measured on impurities with spin fluctuation theory reveals a continuous evolution from negligible impurity bound state features for weaker scattering potential to clearly detectable states for somewhat stronger scattering potentials. All bound states for these intermediate strength potentials are pinned  at or close to the gap edge of the smaller gap, a phenomenon that we explain and ascribe to multi-orbital physics.
\end{abstract}


\maketitle

While there is strong evidence for $s_\pm$ superconductivity in many of the iron-based superconductors, this  may not be universal, and in some of them there are claims that other order parameters prevail \cite{Chubukov2012, reid_universal_2012,Hirschfeld_CRP_16}. LiFeAs is a stoichiometric superconductor ($T_{\mathrm c}$ = 17 K), making it particularly amenable to a comparison with theory. Further, it exhibits an atomically flat non-polar surface that does not undergo reconstruction -- making it suitable for spectroscopic studies of the order parameter by ARPES and STM. Crude aspects of the gap structure of LiFeAs have now been fairly well established by these methods \cite{allan_anisotropic_2012, umezawa_unconventional_2012, borisenko_one-sign_2012,Chi_PRL,chi_sign_2014}.
The overall agreement of multiple experimental groups and methods led to several theoretical attempts\cite{Wang13,Ahnetal14,Yin14,KontaniLiFeAs14} to calculate the detailed gap function, all of which led to the identification of sign-changing s-wave gaps, but differed on the sets of Fermi surface pockets that manifested the same sign. Disputes over these details illustrate the current capabilities of  materials-specific calculations of superconducting properties.  One way to distinguish among these various proposals is to test their predictive power for impurity states, sensitive probes of gap symmetry and structure.

In the past several years, significant progress has been made in realistic simulations of the STM tunneling conductance in superconductors\cite{Markiewicz09,Choubey14,Kreisel15,Demler16}. In the present work, we compare theoretical predictions of conductance spectra for LiFeAs with experiment. While the results may not provide direct information on the origin of Fe-based superconductivity, they are an important indicator of the state of progress towards a quantitative theory of superconductivity in these materials. To this end we study iron-site defects, both engineered by deliberate addition of manganese, cobalt and nickel during the growth of the crystals as well as native defects.

\begin{figure}[ht!]
\includegraphics[width=0.48\textwidth]{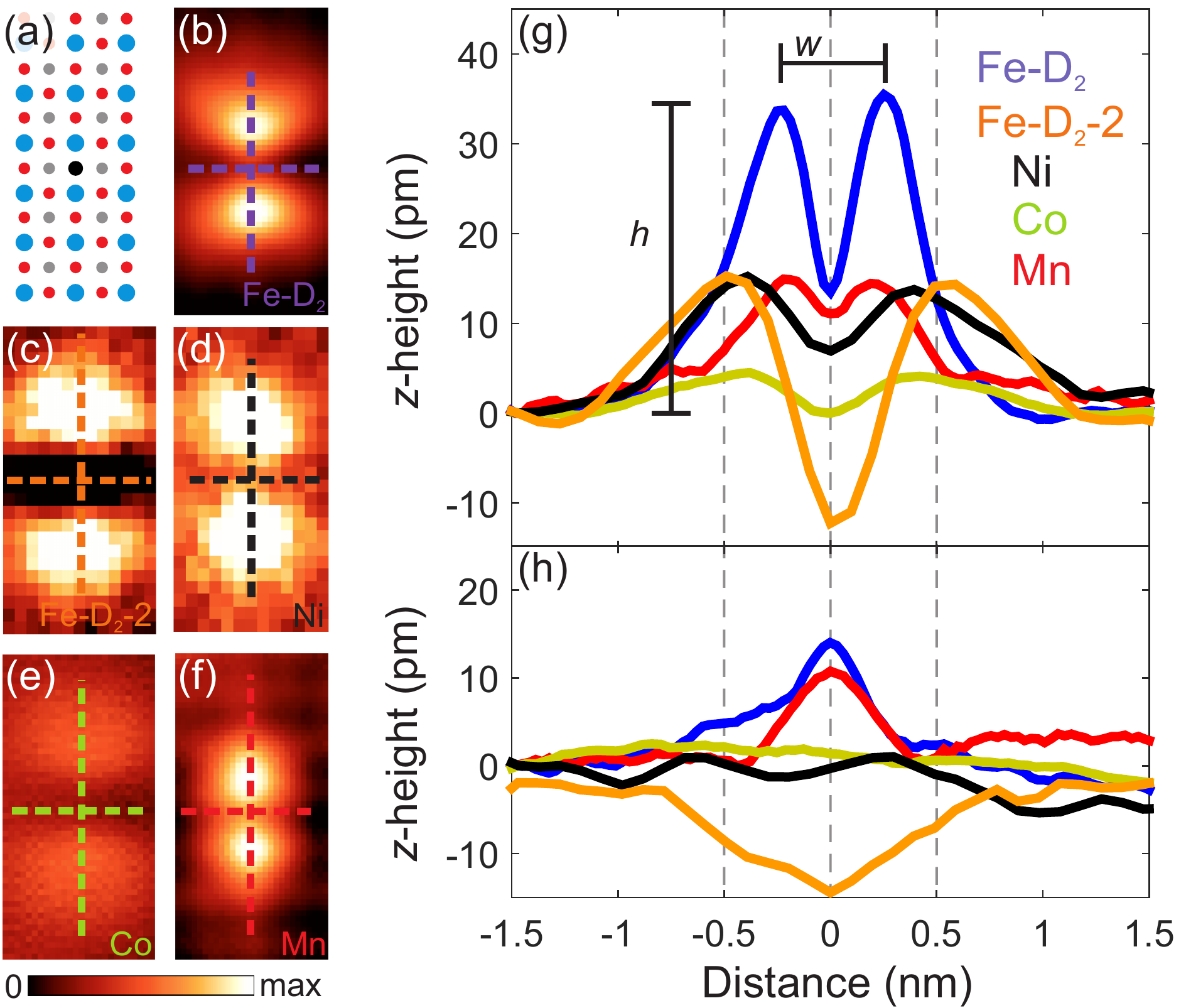}
\caption{\textbf{Appearance of Iron-site defects.} (a) Atomic configuration of the crystal structure of LiFeAs. The defect site is highlighted in black. The gray dots represent Li atoms, red dots represent Fe atoms and blue dots represent As atoms. (b-f) Topographs showing (b) a native defect of type Fe-$D_2$,  (c) a native defect of type Fe-$D_2$-$2$, (d) a single Ni impurity, (e) a Co impurity   and  (f) a Mn impurity. Topographs in (b-f) are imaged at ($V_s=-50\mathrm{mV}$, $50\mathrm{pA}$ (Fe-$D_2$), $50\mathrm{pA}$ (Fe-$D_2$-$2$)), $50\mathrm{pA}$ (Ni), $100\mathrm{pA}$ (Co),  $I=300\mathrm{pA}$ (Mn). (g) Line cuts along the impurities (b-f). The direction of the line cuts along and normal to the impurity are indicated by a dotted lines in (b-f). (h) Line cuts taken normal to the impurities.}
\label{Linecut}
\end{figure}

LiFeAs crystals cleave between the Li layers, exposing a square lattice of lithium atoms at the surface [see Fig.~\ref{Linecut}(a)]. The square lattice seen in STM topography has an orientation and lattice constant compatible with the positions of the As or Li atoms on the surface and is rotated $45^\circ$ with respect to the iron square lattice.
Experiments were performed in a home-built low temperature STM operating at temperatures down to $1.5~\mathrm{K}$ and in magnetic fields up to $14$ $\mathrm T$ in cryogenic vacuum\cite{white_stiff_2011}. The surface was prepared by {\it in-situ} cleaving at low temperatures. We used STM tips cut from a PtIr wire. Bias voltages are applied to the sample, with the tip at virtual ground. Differential conductance spectra have been recorded through a lock-in amplifier with $f=413$ $\mathrm{Hz}$ and a modulation of $V_\mathrm{mod}=500$ $\mathrm{\mu V}$, unless stated otherwise. Data obtained in the superconducting state have been recorded at a temperature of $1.5$ $\mathrm K$.

The defects discussed here are all substituted at the iron site, which has a $D_2$ symmetry. In topographic STM images, the defects share a common overall appearance, which however differs in details [compare Fig.~\ref{Linecut}(b)-(f)] such as the apparent height and the spatial extent. "Engineered" (chemically substituted) defects can be identified from the topographic imaging as they exhibit a different apparent height in comparison to intrinsic defects in LiFeAs \cite{Grothe_PRB} as well as by their concentration. Fig.~\ref{Linecut}(g),(h) shows line cuts taken along the main axis and normal to it through the defects. All images in Fig.~\ref{Linecut} were obtained using the same bias voltage, $-50$ $\mathrm{mV}$. The height with which the defects are seen in topographic images is largest for the native iron-site defects and smallest for cobalt defects, whereas Ni and Mn show similar profiles. Not only the apparent height, but also the separation of the maxima changes for different types of iron-site defects.

\begin{figure}[tb]
\includegraphics[width=0.48\textwidth]{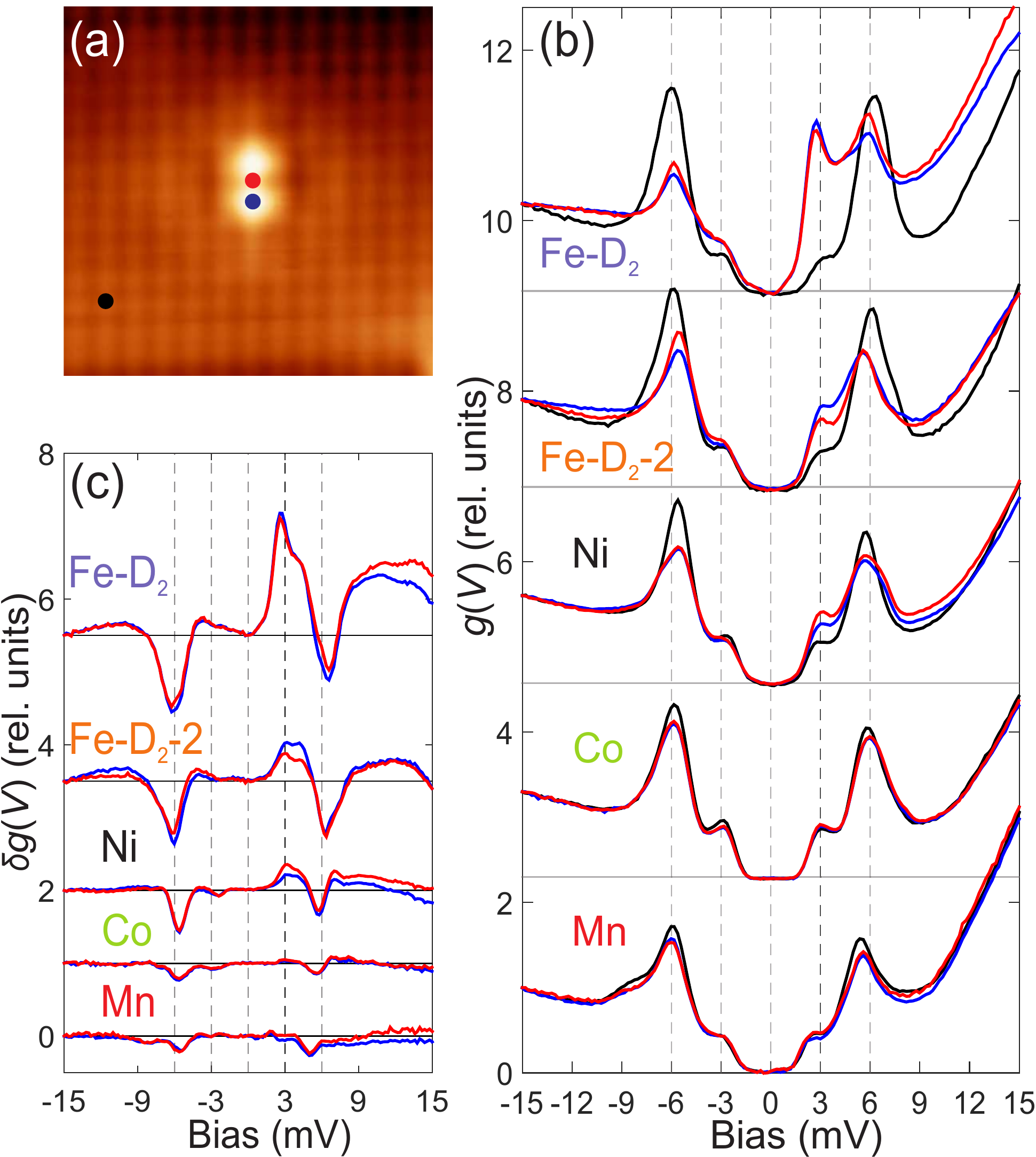}
\caption{ \textbf{Impurity bound states.} (a) Topography showing a single manganese impurity ($V_s=-50\mathrm{mV}$, $I_s=300\mathrm{pA}$, $4.5\times4.5$ nm$^{2}$). (b) Tunneling spectra taken at the impurity site of single Mn, Co, Ni and native iron-site defects. The position of the spectra recorded are color coded as shown in (a). Vertical dashed lines indicate the bias corresponding to $\Delta_1$ at $\pm6$ meV, $\Delta_2$ at $\pm3$ meV, and $E_F$ at $0$ meV. Spectra are normalized at $-15\mathrm{mV}$. (c) Difference $\delta g(V)$ between the impurity spectrum $g_\mathrm{imp}(V)$ and a spectrum obtained with the same tip on the bare surface ($g_0(V)$).}
\label{Spectra}
\end{figure}

Tunneling spectra obtained on both engineered and native iron-site defects are shown in Fig.~\ref{Spectra}. Next to the differential conductance spectra, we also plot the change in differential conductance due to presence of the impurity $\delta g(V)=g_\mathrm{imp}(V)-g_0(V)$.
Of the engineered defects, Mn, Co and Ni, only for Ni is a clear bound state seen. Mn and Co, which both differ by only one electron from iron in the occupation of their $d$-orbital, show only weak changes in the tunneling spectrum.

A slight suppression of the coherence peaks is detected for both, Mn and Co, defects. Subtracted spectra reveal a weak bound state for Co (see Fig.~\ref{Spectra}(c)). The rather weak influence of cobalt defects on superconductivity is consistent with previous STM/STS reports in NaFeAs\cite{yang_unexpected_2012}.

Nickel, which has an almost full $d$ shell, exhibits a stronger scattering potential. Spectra reveal indeed a much stronger bound state at the smaller, positive  bias gap edge [Fig.~\ref{Spectra}(b)] compared to cobalt and manganese. The bound state is strongest at the center between the two lobes and exhibits a pronounced asymmetry between positive and negative bias voltages. Outside the superconducting gap, within $\pm 50$ meV, there is no significant modification of the tunneling spectrum in comparison to the clean surface.

Native iron-site defects occurring in as-grown LiFeAs have been reported previously \cite{Grothe_PRB}. These defects are most likely due to impurities in the source materials, e.g. Si or Al, substitution by other elements of the growth materials, i.e. Li or As, or iron-site vacancies\cite{Huang16}.
Our spectra show a strong bound state resonance near the smaller superconducting gap at positive bias voltages. This bound state has been observed previously. Here, however, due to the higher energy resolution, we can clearly determine its energy, as it is resolved separate from the coherence peak of the larger gap. As already seen to a much weaker extent in the case of Nickel, the defect spectrum again exhibits a pronounced particle-hole asymmetry.

A remarkable feature of all Fe-site defects, which clearly differ substantially in strength of their scattering potential, evidenced from the differing strengths of the bound state features, is that they are all located at roughly the same energy, namely at positive bias close to the position of the smaller gap feature, as indicated by the vertical dashed line at $3$ $\mathrm{meV}$ in Fig.~\ref{Spectra}.  While in a sign-changing superconductor one naively expects a weak nonmagnetic impurity potential to create a bound state just below the gap edge, the features observed here are not located there.  Instead, they are located at the gap edge, in a region of continuum states that one might expect to completely damp such a state.

\begin{figure}[tb]
\includegraphics[width=\linewidth]{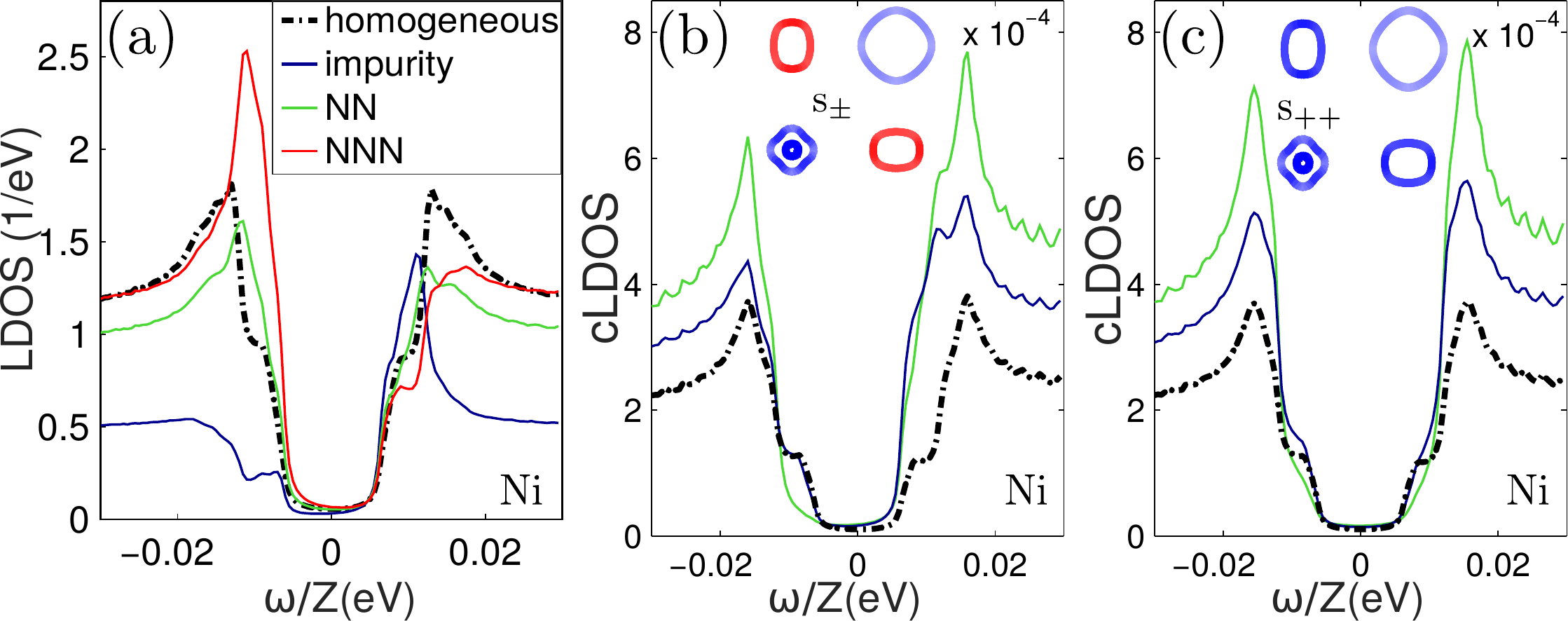}
\caption{(a) Lattice LDOS for an $s_{\pm}$ order parameter  and a Ni impurity.  Note the strong enhancement at negative bias at the NN and NNN positions relative to the impurity.
(b) Continuum LDOS in $[eV\mathrm{bohr}^3]^{-1}$, taking into account the coupling of the electronic states to the tip (``NN" refers here to the cLDOS at the dimer maximum, as in experiment (see Fig. \ref{Spectra}(a)). The largest change in the density of states due to the defect is now seen at positive bias. (inset: calculated gap function on LiFeAs Fermi surface: red = +, blue = -)
(c) The continuum LDOS obtained assuming the same gap magnitude but with equal signs of the order parameter on all Fermi surfaces ($s_{++}$, inset). The change in the LDOS on the impurity is almost symmetric with respect to zero bias.}
\label{fig_spectra_wrong}
\end{figure}

\begin{table}[b!]
\centering
\begin{tabular}{c c c c c c}
 \hline
 Defect & $n_{d}$ &  $E_B$/$\Delta_i$ & Height \emph{h} (pm) & \emph{w} (nm) & $\Delta E$ (eV)\\ [0.5ex]
 \hline\hline
 Mn           & 5 & -        & 14.92  & 0.44 &  0.27 (0.28) \\
 Co           & 7 & 1.0     &  4.50  &  0.74 & -0.32 (-0.35) \\
 Ni           & 8 & 1.06     & 15.21  & 0.78 & -1.76 (-0.87) \\
 Fe-$D_2$-$2$ & - & 1.0     & 15.28  & 0.98 & --  \\
 Fe-$D_2$     & - & 0.92     & 35.46  & 0.50 & --  \\ [1ex]
 \hline
\end{tabular}
\caption{Summary of defects: $n_{d}$ is the $d$-shell configuration of the free impurity atoms if known, $E_\mathrm B$/$\Delta_\mathrm i$ is the bound state energy ($E_{\mathrm B}$) normalized to the inner gap ($\Delta_\mathrm i$) determined from spectra taken on the clean STS ($3\pm0.2\mathrm{mV}$), \emph{h} the apparent height of the defect from STM images obtained at $-50\mathrm{mV}$, and \emph{w} the distance between the maxima (see also Fig.~\ref{Linecut}g). The last column shows the calculated onsite potential (orbitally averaged and unrenormalized) difference at the impurity for LiFeAs and in brackets the values for LaFeAsO as reported in Ref.~\onlinecite{nakamura_first-principles_2011}.}
\label{table:1}
\end{table}

{\it Theory.} We model impurities and the electronic structure of the host on the same footing, as a step towards a quantitative theory of Fe-based superconductivity generally, as well as an attempt to understand which aspects of the superconducting gap structure and symmetry are probed by impurity states in this case. For the calculations we used a 5-orbital tight-binding model, downfolded from a 10-orbital model from Ref.~\onlinecite{Kreisel16}. Imposing an overall renormalization factor $Z=1/2$ due to correlations yields the correct magnitude of the superconducting gap and roughly agrees with observed renormalizations of the electronic structure in the normal state. Transition metal (TM) impurity potentials are obtained using the methods of Ref.~\onlinecite{Berlijn11}. The local density of states in the superconducting state with a spin-fluctuation generated order parameter is calculated using a combination of the T-matrix approximation and the Wannier function approach of Refs.~\onlinecite{Choubey14,Kreisel15, Kreisel16}.
 In summary, we calculate the continuum LDOS (cLDOS) $\rho(\mathbf{r},\omega)$ at the coordinates $\mathbf{r}=(x,y,z)$ of the STM tip such that
the differential conductance in a STM experiment at a given bias voltage $V$ is given by\cite{Tersoff1985}
\begin{equation}
g(\mathbf{r},eV)=\frac{4\pi e}{\hbar} \rho_t(0) |M|^2 \rho(\mathbf{r},eV)=\alpha\rho(\mathbf{r},eV) \,,
\label{eq_conductance}
\end{equation}
where all terms independent of the position and energy have been collected into the constant $\alpha$.

The orbitally averaged potentials obtained from our {\it ab initio} studies for Ni, Mn, Co are summarized, together with the experimental results, in Table~\ref{table:1} and are very similar in magnitude to those calculated earlier for a different compound\cite{nakamura_first-principles_2011}. The orbital dependence of the impurity potentials is not very pronounced, but has been taken into account for the following calculation.
We start with a discussion of the results for a Ni impurity where the on-site impurity potential, like all potentials in this study, has been renormalized downward by a factor $2.5$.
First, we note that a lattice calculation does not reflect the correct spatial symmetry of the LDOS at the surface\cite{Choubey14} and hence cannot be expected to yield a description of the tunneling spectra. The calculation of the continuum LDOS based on including the Wannier functions correctly reflects the local environment of the atoms at the surface and naturally yields a dimer structure in the direction of the NN As atoms on the surface\cite{Choubey14}.

\begin{figure}[tb]
\includegraphics[width=\linewidth]{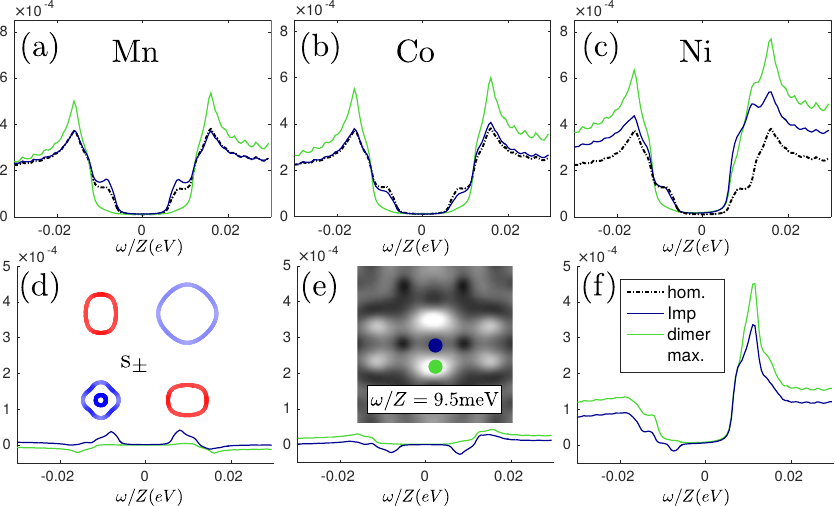}
\caption{Continuum LDOS (a) close to a Mn, (b) Co, and (c) Ni impurity and the corresponding difference spectra (d-f) calculated for the sign-changing order parameter which is shown in the inset of (d). The inset in (e) shows a map of continuum LDOS close to a Ni impurity on an area of $1\;\text{nm}\times1\;\text{nm}$ together with the positions of the spectra. All impurity potentials divided by 2.5 relative to {\it ab initio} calculations.}
\label{fig_spectra_cont}
\end{figure}

While for a lattice calculation, the Ni impurity has largest effect at negative bias,  see Fig. \ref{fig_spectra_wrong}(a), in the continuum LDOS shown in Fig.~\ref{fig_spectra_wrong}(b) we see that the bound states occur at positive bias independent of the relative position to the impurity, in agreement with experiment.
This improved treatment  of particle-hole weights was noted earlier\cite{Kreisel15} and is generally due to a cancellation of NN and NNN Wannier functions.
In contrast, a calculation where the order parameter is forced to have identical magnitudes but positive sign on all bands (Fig.~\ref{fig_spectra_wrong}(c)) shows no impurity bound states, as expected.
Results for  continuum  spectra for all three TM impurities are compared in Fig.~\ref{fig_spectra_cont}.  In each case, the {\it ab initio} impurity potentials have been divided by a factor of 2.5. As in experiment, there is little influence from Co and Mn impurities, whereas for Ni, the impurity resonance is located primarily on the positive bias side, at the position of the lower gap edge\cite{gastiasoro_impurity_2013}.

\label{sec:discussion}
{\it Discussion.} It is well known that while {\it ab initio} approaches capture qualitative features of the electronic structure of Fe-SC, they cannot accurately describe many important near-Fermi level properties of Fe-SC in general\cite{Biermann16} and LiFeAs in particular\cite{Ferberetal12,Haule12}. These uncertainties represent some of the underlying reasons why the pairing state in this material is still under debate. The results of our theoretical calculations, derived from DFT-based band structures, agree quite well in many qualitative respects with the experimental findings, but clearly do not represent a complete solution to the problem.  In the following, we list discrepancies and possible explanations together with proposals to investigate them more deeply.

The impurity potentials obtained from a first principles calculation are too strong in magnitude to yield the weak in-gap bound states at the lower gap edge observed experimentally. Since the identities of the engineered impurities are well known, and they are well isolated, a one-impurity problem for the given chemical substituent is appropriate.  It seems likely therefore that the {\it ab initio} method simply overestimates the potentials due to neglect of correlations, which are known to be significant\cite{Ferberetal12,Haule12}. Assuming a screening that is comparable for all TM impurities, the impurity potentials were multiplied with the same overall renormalization, ultimately  giving  a reasonable agreement of the spectra between experiment and theory. Thus the relative strengths and the signs of the potentials seem to be calculated correctly.

One interesting question is why the bound state energies occur at the smaller gap, and in fact augment the LDOS there. As noted above, in a one-band system, weak potentials produce bound states just {\it below} the gap edge, and are essentially invisible. A realistic impurity in a multi-orbital system is approximately diagonal in orbital space, and in this case the condition for a bound state in a sign-changing $s$-wave state may be shown to decouple into distinct orbital channels,
\begin{equation}
0\approx 1-V^{\mu\mu}_{\text{imp}} G^0_{{\bf R}=0}(\omega)^{\mu\mu},
\end{equation}
for $\mu$ equal to any of the five $d$-orbitals, and $ G^0_{{\bf R}=0}(\omega)^{\mu\mu}$ the local diagonal Nambu Green's function corresponding to orbital $\mu$.
A resonant effect of a bound state in, e.g. the $d_{yz}$ channel on the measured LDOS  can appear, however, only if the damping is sufficiently small, i.e.  when the orbitally resolved LDOS  in this channel vanishes.  This occurs below the (larger) $d_{yz}$ gap.  The continuum states that exist below this energy, i.e. the $d_{xy}$ states,  can broaden this bound state only to the extent the two orbitals mix, due to relatively weak band hybridization and spin-orbit coupling effects.  The net result is that the impurity resonance appears to enhance the lower gap coherence peak, as seen also in Ref.~\onlinecite{gastiasoro_impurity_2013}.

While this property is a consequence of the structure of the local lattice Green function, the relative magnitude of
the bound states at positive and negative energies as measured in an STM experiment can only
be described properly using the continuum LDOS  following Eq. (\ref{eq_conductance}). In the present case, the relative particle-hole weights
are switched when doing the basis transformation, giving the primary spectral weight on the dimer maximum at positive bias, in agreement with experimental findings.
The spectra presented in Figs.~\ref{fig_spectra_wrong} (b,c) and \ref{fig_spectra_cont} are calculated
at a fixed tip height. Due to the overall differences in the magnitude of the continuum LDOS close
to the impurity, the spectra are enhanced at larger energies, in contrast to what is seen in the experimental conductance spectra.
A more realistic calculation would take into account the size and local wavefunction of the STM tip, which would give a more homogeneous conductance spectrum at these higher energies to the extent it samples more points within the unit cell. We have not investigated these effects here because it is difficult to estimate the microscopic details of the STM tip.

In conclusion, we have presented STM measurements and a detailed theoretical analysis of Fe-site defect states in the Fe-based superconductor LiFeAs, with the intention of distilling what can be learned about the superconducting state in this material, and refining techniques for future analysis.
The theoretical analysis finds good qualitative agreement with  local spectra and impurity bound states, but only if the {\it ab initio} impurity potentials are renormalized, confirming earlier suggestions that correlations play a significant role in the electronic structure. The tendency of the bound states to occur at a fixed energy at the lower coherence peak position over a range of impurity potentials is explained by the tendency of weak potentials to form bound states first just below the large gap, together with an approximately orbitally diagonal impurity potential that allows the bound state to remain sharp only in an energy range where other continuum states couple weakly to the bound state due to their differing orbital character. In consequence, the bound state is noticeable only close to the energy of the smaller gap. Of course, the existence of these peaks requires an $s_\pm$ state, as we explicitly showed by calculation of spectra assuming an analogous $s_{++}$ state for comparison. The good agreement of theory and experiment reported here bodes well for a future true quantitative analysis of inhomogeneous superconductivity and STS spectra in these systems.

\begin{acknowledgments} The authors acknowledge useful discussions with C. Hess, Y. Wang and D. Guterding.
SC, DB and PW acknowledge funding from the MPG-UBC center. PW acknowledges financial support from EPSRC (EP/I031014/1). PJH was supported by NSF-DMR-1407502.
A portion of this research was conducted at the Center for Nanophase Materials
Sciences, which is a Department of Energy (DOE) Office of Science User Facility.
\end{acknowledgments}

\bibliographystyle{apsrev4-1}
\label{Bibliography}
\bibliography{lifeas_eng}
\end{document}